\documentclass[pre,showpacs,superscriptaddress,floatfix]{revtex4}
\usepackage[latin1]{inputenc}
\usepackage{graphicx}
\usepackage{amsmath}
\usepackage{amssymb}
\usepackage{color}
\begin{document}

\title{Comment on ``Drainage of a Thin Liquid Film between Hydrophobic Spheres:
Boundary Curvature Effects''}

\author{Evgeny S. Asmolov}
\affiliation{A.N.~Frumkin Institute of Physical Chemistry and Electrochemistry, Russian Academy of Sciences, 31 Leninsky
Prospect, 119071 Moscow, Russia}
\affiliation{Central Aero-Hydrodynamic Institute, 140180
Zhukovsky, Moscow region,  Russia}
%\alsoaffiliation{Institute of Mechanics, M. V. Lomonosov Moscow State University, 119991 Moscow,
%Russia}

\author{Olga I. Vinogradova}
\affiliation{A.N.~Frumkin Institute of Physical Chemistry and Electrochemistry, Russian Academy of Sciences, 31 Leninsky
Prospect, 119071 Moscow, Russia}
\affiliation{DWI - Leibniz Institute for Interactive Materials, RWTH Aachen, Forckenbeckstr. 50, 52056 Aachen, Germany}
\affiliation{Department of Physics, M.V.~Lomonosov Moscow State University, 119991 Moscow, Russia}
\email{oivinograd@yahoo.com}

%\begin{tocentry}

%  \includegraphics[width=2.5cm]{Asmolov_Fig1.eps}\\

%\end{tocentry}

\begin{abstract}

  We clarify the basic faults in Fang, A.; Mi, Y. Drainage of a Thin Liquid Film between Hydrophobic Spheres: Boundary
Curvature Effects. Langmuir 2014, 30, 83-89, which led to unphysical conclusions.

\end{abstract}
\maketitle

\newpage

 In their paper~\cite{fang.a:2014} Fang and Mi declare that by including the \textsl{boundary curvature effect} into the hydrodynamic slip boundary conditions, they reformulate and improve the Vinogradova theory of a thin film drainage between hydrophobic surfaces.~\cite{vinogradova.oi:1995a} As a side note, we remark that results~\cite{vinogradova.oi:1995a} are general, being applicable, besides hydrophobic surfaces in water,~\cite{vinogradova:03,charlaix.e:2005} to a description of a drainage of confined polymers,~\cite{horn:00,honig.cdf:2008}  gases~\cite{maali.a:2008,seo.d:2013}, or any other systems characterized by a partial slip length, $b$.

   The solution by Fang and Mi significantly underestimates the force and does not recover the classical Taylor (Reynolds) result~\cite{chan.dyc:1985} expected at large distances, $h \gg b$. Besides that, a curvature contribution to the slip length~\cite{einzel1990}, underlying the boundary conditions~\cite{fang.a:2014}, should obviously become important only  for spheres of a small radius, but not with numerical parameters used by Fang and Mi.

  In the present Comment,  we clarify the mistakes by Fang and Mi,~\cite{fang.a:2014} and point out that the correction to the Vinogradova theory due to an extra curvature term cannot be derived within the lubrication approximation.

   \begin{figure}[tbp]
\includegraphics[width=0.3\textwidth]{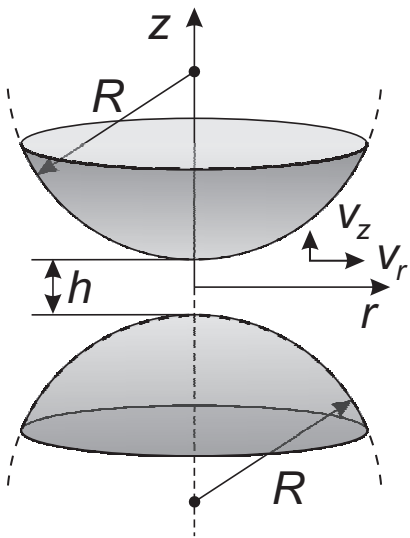}
\caption{Definition sketch for two  approaching (with the velocity $V$) spheres of radius $R$ separated by the gap $h = \varepsilon R$, where $\varepsilon\ll 1$.  Cylindrical coordinates $(r,z)$ are used with origin in the center of the gap, so that the equations of the sphere surfaces become $\displaystyle \pm H (r) \simeq \pm\frac{h}{2} \pm \frac{r^2}{2 R}$. The radial and normal velocity components are $v_r$ and $v_z$.}
\label{sketch}
\end{figure}

There are at least three basic faults in Ref.~\cite{fang.a:2014} which lead to erroneous conclusions. To highlight them we focus on the case of two identical spheres studied analytically by Fang and Mi (see Fig.~\ref{sketch}).

\emph{First}, the lubrication approximation, Eqs. (12) and (13) by Fang and Mi, implies that $\varepsilon \ll 1$, so that
$v_{r} = O( \varepsilon ^{-1/2}V)$ and $r = O(\varepsilon ^{1/2}
R)$. In such a case in the set of Eqs.(10) and (11) left ones are fully identical to formulated by Vinogradova,~\cite{vinogradova.oi:1995a} and include the boundary curvature term, $\displaystyle \frac{rv_{r}}{R} = O(V)$. Additional terms due to curvature, $\displaystyle\frac{r V}{R} = O(\varepsilon ^{1/2} V)$, included by Fang and Mi into right equations of (10) and (11), should be neglected in the leading-order solution being out-of-order. These could only be included if terms of the same order are kept in Eqs. (5), (6), (8) by Fang and Mi. To construct then  the second-order corrections to pressure and velocity fields it is necessary to match the inner solution in the gap with the outer solution
at distances $r = O(R).$ Such a problem is certainly beyond the scope of the lubrication approximation.

\emph{Second,} integrating the continuity equation, Eq.(7) by Fang and Mi, over $z$ should be made as
follows~\cite{vinogradova:96}:%
\begin{eqnarray}
\int_{-H\left( r\right) }^{H\left( r\right) }\left[ \frac{\partial v_{z}}{%
\partial z}+\frac{1}{r}\frac{\partial }{\partial r}\left( rv_{r}\right) %
\right] dz&=&v_{z}\left( H\right) -v_{z}\left( -H\right)\nonumber\\
&+&\frac{1}{r}\frac{d}{%
dr} \int_{-H\left( r\right) }^{H\left( r\right) }rv_{r}dz-\frac{dH}{dr}%
\left( v_{r}\left( H\right) +v_{r}\left( -H\right) \right) =0.\nonumber
\end{eqnarray}%
By substituting here $v_{z}\left( H\right) $ and $v_{z}\left( -H\right) $
from their Eqs.(10) and (11) we then obtain%
\begin{equation}
\frac{r}{R}\left[ v_{r}\left( H\right) +v_{r}\left( -H\right) \right] -V+%
\frac{1}{r}\frac{d}{dr}\left( \int_{-H\left( r\right) }^{H\left( r\right)
}rv_{r}dz\right) -\frac{dH}{dr}\left[ v_{r}\left( H\right) +v_{r}\left(
-H\right) \right] =0,  \label{2}
\end{equation}
and conclude that Fang and Mi missed the
last term in our Eq.(\ref{2}). Indeed, the first and last terms are obviously canceled out since $\displaystyle \frac{dH}{dr}\simeq\frac{r}{R},$
and we get%
\begin{equation}
-V+\frac{1}{r}\frac{d}{dr}\left( \int_{-H\left( r\right) }^{H\left(
r\right) }rv_{r}dz\right)
=-V+\frac{1}{r}\frac{d}{dr}\left( -\frac{H^{3}}{3\mu }\frac{dp}{dr}%
+2C_{2}H\right) =0.
\end{equation}%
Thus, constant $C_{1}$ in the velocity profile, Eq.(14) by Fang and Mi, which they claim to represent a ``curvature-induced renormalization of the slip length'', does not contribute
to the integral of the continuity equation.

\emph{Third}, Eq.(15) derived by Fang and Mi gives a
non-zero value for constant $C_1$, which implies that their velocity profile is asymmetric. Since the problem for two equal spheres is symmetric, $v_{r}$ is necessarily an
even function of $z,$ which requires $C_{1}=0.$ This error is likely
caused by a wrong sign for one of the boundary curvature terms, $\displaystyle \frac{r V}{R}$, in their Eqs. (10) and (11).

In summary, although the extension of the Vinogradova theory~\cite{vinogradova.oi:1995a} to the case of highly curved surfaces would be very timely, the theory by Fang and Mi~\cite{fang.a:2014} fails in trying this.

\bibliographystyle{unsrt}
%\bibliography{lcomment_bib}

\providecommand{\latin}[1]{#1}
\providecommand*\mcitethebibliography{\thebibliography}
\csname @ifundefined\endcsname{endmcitethebibliography}
  {\let\endmcitethebibliography\endthebibliography}{}

%your .bib file

\end{document}